\documentstyle[aps,prl,epsf,floats]{revtex} 
\begin{document}
\twocolumn[\hsize\textwidth\columnwidth\hsize\csname@twocolumnfalse%
\endcsname

\title{Damage spreading and dynamic stability of kinetic Ising models}
\author{Thomas Vojta}
\address{Department of Physics and Materials Science Institute,
University of Oregon,Eugene, OR 97403 \\
and Institut f{\"u}r Physik, TU Chemnitz-Zwickau,
D-09107 Chemnitz, Germany.}

\date{\today}
\maketitle

\begin{abstract}
We investigate how the time evolution of different kinetic Ising models
depends on the initial conditions of the dynamics. To this end we consider
the simultaneous evolution of two identical systems subjected to the
same thermal noise. We derive
a master equation for the time evolution of a joint probability 
distribution of the two systems. This equation is then solved within
an effective-field approach. By analyzing the fixed points of the master
equation and their stability we identify regular and chaotic phases. 
\end{abstract}
\pacs{PACS numbers: 05.40.+j, 64.60.Ht, 75.40.Gb} 

]  

 
The question to what extent the time evolution of a physical system
depends on its initial conditions is one of the central questions 
in nonlinear dynamics that have lead to the discovery of chaotic 
behavior \cite{chaos}. In more recent years analogous concepts have
been applied to the stochastic time evolution of interacting 
systems with a macroscopic number of degrees of freedom.
Among the simplest of such many-body systems are kinetic Ising models where
the above question has been investigated by means of so-called
"damage spreading" simulations \cite{stanley,derrida}. In these Monte-Carlo 
simulations two identical systems with different initial conditions 
are subjected to the same thermal noise, i.e. the same random numbers 
are used in the Monte-Carlo procedure. The differences in the microscopic
configurations of the two systems are then used to characterize the 
dynamic stability.

Later the name "damage spreading" has also been applied to a different
though related type of investigations in which the two systems are {\em not}
identical but differ in that one or several spins in 
one of the copies are permanently fixed in one direction. Therefore the
equilibrium properties of the two systems are different and
the microscopic differences between the two copies can be related to
certain thermodynamic quantities \cite{coniglio,glotzer}. Note that in 
this type
of simulations the use of identical noise (i.e. random numbers) for the
two systems is not
essential but only a convenient method to reduce the statistical error.
Whereas this second type of damage spreading is well understood and 
established as a method to numerically calculate equilibrium correlation
functions, much less is known about the original problem of dynamic
stability. In particular, the transition between regular and chaotic 
behavior (called the "spreading transition") is not understood and it 
is unknown under which conditions it coincides 
with the equilibrium phase transition of the Ising model. (In numerical
simulations it seems to depend on the dynamical algorithm whether the
two transitions coincide or not. Glauber dynamics usually gives a 
spreading temperature which is slightly lower than the equilibrium 
critical temperature whereas the two are identical for heat-bath 
dynamics \cite{ts}.)

In this Letter we therefore concentrate on the original question of
the stability of the stochastic dynamics in kinetic Ising models.
To this end we investigate the time evolution of two {\em identical}
systems with different initial conditions which are subjected to the same
thermal noise. We derive a master equation for the joint
probability distribution of the two systems and solve it within
an effective-field approach. By analyzing the fixed 
points of this equation we identify regular and chaotic phases. 
We find that the location of these phases in the  phase diagram is
sensitive to the choice of the dynamic algorithm. In particular,
Glauber dynamics and heat-bath dynamics give very different dynamical
phase diagrams.

We consider two identical kinetic Ising models with $N$ sites, 
described by the Hamiltonian
\begin{equation}
H = - \frac 1 2  \sum_{ij} J_{ij} S_i S_j - h \sum_i S_i
\end{equation}
where $S_i$ is an Ising variable with the values $\pm 1$. The dynamics 
is given by one of the two following stochastic maps which describe
Glauber dynamics 
\begin{mathletters}
\begin{equation}
S_i (t+1) = {\rm sign} \left\{ v[h_i(t)] 
- \frac 1 2 +S_i(t) \biggl[ \xi_i(t) - \frac 1 2 \biggr] \right\}
\end{equation}
and heat-bath dynamics
\begin{equation}
S_i (t+1) = {\rm sign} \left\{ v[h_i(t)] - \xi_i(t) \right\}
\end{equation}
\end{mathletters}
with
\begin{equation}
v(h) = {e^{h/T}/ ({e^{h/T}+ e^{-h/T}}}).
\end{equation}
Here $h_i(t)=\sum_j J_{ij} S_j(t) + h$ 
is the local magnetic field at site $i$ and (discretized)
time $t$, $\xi_i(t) \in [0,1)$ is a random number which is identical
for both systems, and $T$ denotes the temperature. Note that Glauber-
and heat-bath algorithm differ only in the way how the random numbers 
are used to update the configuration. The transition {\em probabilities}
$v$ are identical for both algorithms.

In order to describe the simultaneous time evolution of two systems
$H^{(1)}$ and $H^{(2)}$ with Ising spins $S_i^{(1)}$ and $S_i^{(2)}$ 
we define a variable $\nu_i(t)$ with the values $\nu = ++$ for 
$S^{(1)}=S^{(2)}=1$, $+-$ for $S^{(1)}=-S^{(2)}=1$, $-+$ 
for $-S^{(1)}=S^{(2)}=1$  and $--$ for $S^{(1)}=S^{(2)}=-1$
which describes the state of a spin pair $(S^{(1)},S^{(2)})$. Since we are
interested in the time evolution not for a single sequence of $\xi_i(t)$,
but in $\xi$-averaged quantities we consider a whole ensemble of 
system pairs $(H^{(1)}$,$H^{(2)})$ and define a probability distribution 
\begin{equation}
P(\nu_1,\ldots,\nu_N,t) = \left \langle \sum_{\nu_i(t)} \prod_i 
\delta_{\nu_i,\nu_i(t)} \right \rangle
\end{equation}
where $\langle \cdot \rangle$ denotes the ensemble average.
The time evolution of $P(\nu_1, ...\nu_N,t)$ for a single-spin dynamical
algorithm as, e.g., Glauber or heat-bath dynamics is given 
by a master equation 
\begin{eqnarray}
\lefteqn{\frac d {dt} P(\nu_1,\ldots,\nu_N,t) =} \nonumber\\
& &- \sum_{i=1}^N \sum_{\mu_i \not= \nu_i}
P(\nu_1,\ldots,\nu_i,\ldots,\nu_N,t) w(\nu_i \to \mu_i) \nonumber\\
& &+ \sum_{i=1}^N \sum_{\mu_i \not= \nu_i}
P(\nu_1,\ldots,\mu_i,\ldots,\nu_N,t) w(\mu_i \to \nu_i).
\end{eqnarray}
Here $w(\mu_i \to \nu_i)$ is the transition probability of the spin-pair
$(S_i^{(1)},S_i^{(2)})$ from state $\mu$ to $\nu$. It is a function of the
local magnetic fields $h_i^{(1)}$ and $h_i^{(2)}$ and can be calculated 
from (2a) or (2b) for Glauber and heat-bath dynamics, respectively. 

A complete solution of the master equation (5) is, of course, out of 
question. Therefore one has to resort to approximation methods, the most 
obvious being mean-field approximations. A natural way to construct a
mean-field theory is usually to take the range of the interaction $J_{ij}$
to infinity. However, a mean-field theory constructed this way does not
show the chaotic behavior found in the Glauber Ising model at high
temperatures. A more detailed analysis \cite{vojta} shows that the
absence of any fluctuations in the infinite-range model is responsible
for this discrepancy since the fluctuations are 
essential for the chaotic behavior \cite{derridamf}. 

We therefore develop a slightly more sophisticated 
effective-field approximation that 
retains the fluctuations, though in a quite simplistic manner. 
The central idea is to treat the fluctuations at different sites as
statistically independent. This amounts to approximating the 
probability distribution $P(\nu_1, ...\nu_N,t)$ by a product of 
identical single-site distributions $P_{\nu}$,
\begin{equation}
P(\nu_1,\ldots,\nu_N,t) = \prod_{i=1}^N P_{\nu_i}(t).
\end{equation}
Using this, the master equation (5) reduces to an equation of motion for
the single-site distribution $P_{\nu}$,
\begin{equation}
\frac d {dt} P_{\nu} =  \sum_{\mu \not= \nu} [- P_{\nu} W(\nu \to \mu)
+ P_{\mu} W(\mu \to \nu)],
\end{equation}
where 
\begin{equation}
W(\mu \to \nu) = \left \langle w(\mu \to \nu) \right \rangle_P
\end{equation}
is the transition probability averaged over the states $\nu_i$ of all sites according 
to the distribution $P_{\nu}$. Note that the average magnetizations $m^{(1)}$, $m^{(2)}$
of the two systems and the Hamming distance (also called the damage)
\begin{equation}
D = \frac 1 {2N} \sum_{i=1}^N |S_1^{(1)} - S_i^{(2)} |
\end{equation}
which measures the distance between the two systems in phase space can be
easily expressed in terms of $P$,
\begin{mathletters}
\begin{eqnarray}
m^{(1)} &=& P_{++} + P_{+-} - P_{-+} - P_{--},\\
m^{(2)} &=& P_{++} - P_{+-} + P_{-+} - P_{--}, \\
D &=& P_{+-} + P_{-+}.
\end{eqnarray}
\end{mathletters}

So far the considerations have been rather general, to be specific we will
now concentrate on a two-dimensional system on a hexagonal lattice
with a nearest neighbor interaction of strength $J$. The external magnetic
field $h$ is set to zero. To solve the 
master equation (7) for the single-site distribution $P$ we first calculate
the transition probabilities $w(\mu \to \nu)$ between the states
of a spin pair from one of the stochastic maps
(2a) or (2b) and then
average these probabilities over the states of the three neighboring sites 
of a certain reference site with respect to the yet 
unknown distribution $P$. This yields the transfer rates $W(\mu \to \nu)$ 
which enter (7).
The calculations involved are quite tedious but straight forward, they will
be presented in some detail elsewhere \cite{vojta}.

The resulting system of non-linear equations for the variables 
$P_{++}$, $P_{+-}$, $P_{-+}$, and $P_{--}$ can first be used to calculate
the thermodynamics. As expected, Glauber and heat-bath dynamics
give the same results. In particular, there is a ferromagnetic phase 
transition at a temperature $T_c$ determined by
\begin{equation}
 \tanh \frac {3J} {T_c} + \tanh \frac J T_c = \frac 4 3,
\end{equation}
which gives $T_c/J \approx 2.11$. In the ferromagnetic phase
the magnetization is given by
\begin{equation}
m^2 = \frac { \frac 3 4 (\tanh 3J/T + \tanh J/T) -1}  { \frac 3 4 \tanh J/T 
- \frac 1 4 \tanh 3J/T}.
\end{equation}

We now discuss the time evolution of the Hamming distance $D$ between the
the two systems which characterizes the stability of the dynamics. In 
contrast to the thermodynamics Glauber and heat-bath algorithms give very
different results for the Hamming distance. We first consider the Glauber case.

The equation of motion of the Hamming distance can easily be derived from
(7) and (10c). In the paramagnetic phase we obtain after some algebra
\begin{equation}
\frac d {dt} D = \frac 1 2 (D - 3 D^2 + 2 D^3) \tanh \frac {3J} T .
\end{equation}
This equation has three stationary solutions, i.e. fixed points, $D^*$, 
viz. $D_1^*=0$
which corresponds to the two systems being identical, $D_2^*=1$ where
$S^{(1)}=-S^{(2)}$ for all sites, and $D_3^*=1/2$ which corresponds to
completely uncorrelated configurations. To investigate the stability
of these fixed points we linearize (13) in $d=D-D^*$. The linearized equation  
has a solution $d \propto e^{-\lambda t}$ with
$\lambda_1 =\lambda_2 = \frac 1 2 \tanh 3J/T$ and $\lambda_3 = - \frac 1 4 
\tanh 3J/T$. Consequently, the only stable fixed point is $D_3^*= 1/2$.
Thus, in the paramagnetic phase the Glauber dynamics is chaotic, since
two systems, starting close together in phase space ($D$ small initially) 
will become separated exponentially fast with a Lyapunov exponent 
$\lambda_1$, eventually reaching a stationary state with an
asymptotic Hamming distance $D=1/2$. Note, that
the Lyapunov exponent $\lambda_1$ goes to zero for $T \to \infty$. Therefore
the time it takes the systems to reach the stationary state diverges with
$T \to \infty$, as has also been found in simulations \cite{wappler}.

We now turn to the ferromagnetic phase. In order to find the fixed points
of the master equation (7) we can set the magnetizations of both systems 
to their equilibrium values (12) from the outset. In doing so we exclude, 
however, all phenomena connected
with the behavior after a quench from high temperatures to temperatures
below $T_c$. These phenomena require an investigation of the {\em early}
time behavior and will be analyzed elsewhere \cite{vojta}.
For $m^{(1)} = m^{(2)} = m$ the equation of motion for the
Hamming distance reads
\begin{eqnarray}
\lefteqn {\frac d {dt} D = \frac 1 2 (D - 3 D^2 + 2 D^3) \tanh 
\frac {3J} T} \nonumber \\
& & - \frac 3 4 m^2 \left(2 D \tanh \frac J T -D^2 \tanh \frac J T +D^2 \tanh 
\frac {3J} T \right).
\end{eqnarray}
This equation has two fixed points $D^*$ in the interval [0,1]. The first,
$D_1^*=0$ exists for all temperatures. The second fixed point $D_3^*$
with $0<D_3^*< \frac 1 2$ exists only for $T>T_s$ where the spreading 
temperature $T_s$ is determined by
\begin{equation}
3 m^2 \tanh \frac J {T_s} =\tanh \frac {3J} {T_s}.
\end{equation}
This gives $T_s \approx 1.74  \approx 0.82 T_c$. The stability analysis shows 
that $D_1^*=0$ is stable for $T<T_s$ and unstable for $T>T_s$ with a 
Lyapunov exponent $\lambda_1 = \frac 1 2 \tanh 3J/T - \frac 3 2 m^2 \tanh J/T$. 
The fixed point $D_3^*$ which exists only for $T>T_s$ is always stable.
Consequently, we find that the Glauber dynamics is regular with the
asymptotic Hamming distance being zero for temperatures
smaller than the spreading temperature $T_s$
but chaotic for $T>T_s$. Close to the spreading temperature the
asymptotic Hamming distance increases linearly with $T-T_s$ which
corresponds to the spreading transition being of 2nd order.
In contrast to the paramagnetic phase, where the two systems become
eventually completely uncorrelated, for $T_s<T<T_c$ the asymptotic
Hamming distance $D$ is always smaller than 1/2 so that the two systems
remain partially correlated. 
Directly at the spreading point the term linear in $D$ in (14) vanishes.
For small Hamming distances the equation of motion now reads
$d D /dt \propto -D^2$ which 
gives a power-law behavior $D(t) \propto t^{-1}$.

Analogously, for $m^{(1)} = -m^{(2)} = m$ we find two fixed points,
$D_2^*=1$ which exists for all temperatures and $D_4^*$ with
$\frac 1 2 < D_4^* <1$ which exists for $T>T_s$ only. $D_2^*$ is stable for 
temperatures $T<T_s$ and unstable for $T>T_s$ whereas $D_4^*$ is always 
stable if it exists.
The results for damage spreading in the Glauber Ising model within our
effective-field approximation are summarized in Fig. 1.
\begin{figure}
  \epsfxsize=7.0cm
  \epsfysize=7.0cm
  \centerline{\epsffile{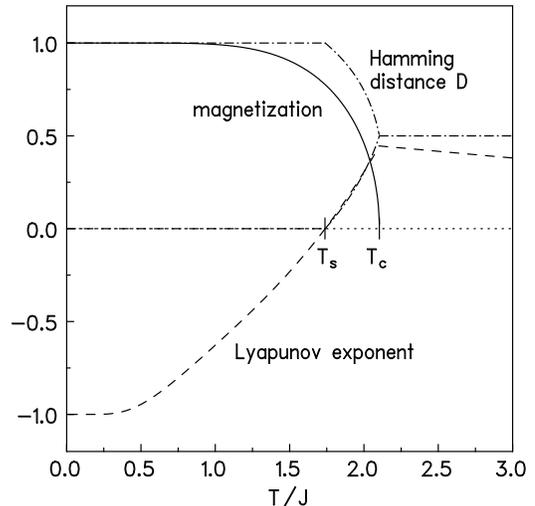}}
  \caption{Magnetization $m$, asymptotic Hamming distance $D$ and Lyapunov
   exponent $\lambda_1$ as functions of temperature for the Glauber Ising
   model. Below $T_c$ the curve for $D$ has two branches corresponding to
   the two systems being in the same or in different free energy valleys.}
  \label{fig:1}
\end{figure}

We now investigate the time evolution of the damage for the Ising model
with heat-bath  dynamics (2b). After calculating the averaged transition
rates $W(\mu \to \nu)$ \cite{vojta} and inserting them into (7), we
obtain the equation of motion for the Hamming distance $D$. In the
paramagnetic phase it reads
\begin{eqnarray}
\lefteqn{ \frac d {dt} D = \frac {3D} 4  \Bigl[ \tanh \frac {3J} T + 
  \tanh \frac J T - 4/3  \Bigr] }  \\
& & - \frac {3 D^2} 4 \Bigl[ \tanh \frac {3J} T +  \tanh \frac J T  \Bigr]
+\frac {D^3} 4 \Bigl[ \tanh \frac {3J} T +  3  \tanh \frac J T \Bigr]. \nonumber
\end{eqnarray}
This equation has only a single fixed point in the physical interval
[0,1], viz. $D_1^*=0$ \cite{sym}. It is stable everywhere in the paramagnetic phase.
Consequently, the asymptotic Hamming distance is zero for all initial
conditions, and the heat-bath Ising model does not show chaotic behavior for
$T>T_c$. The Lyapunov exponent is given by $\lambda_1= 
\frac 3 4 \tanh \frac {3J} T + \frac 3 4  \tanh \frac J T -1 < 0$. 
The Lyapunov exponent goes to zero for $T \to T_c$, thus at the 
critical temperature we again find for small Hamming distances
$d D /dt \propto -D^2$ which gives $D(t) \propto t^{-1}$.

In the ferromagnetic phase for $m^{(1)} = m^{(2)} =m$ the equation of motion
is given by
\begin{eqnarray}
\lefteqn{ \frac d {dt} D = } \\
& & \frac {3D} 4  \Bigl[ (1+m^2) \tanh \frac {3J} T + 
  (1-3m^2) \tanh \frac J T - 4/3  \Bigr]  \nonumber \\
&- &  \frac {3 D^2} 4 \Bigl[ \tanh \frac {3J} T +  \tanh \frac J T  \Bigr]
+\frac {D^3} 4 \Bigl[ \tanh \frac {3J} T +  3  \tanh \frac J T \Bigr]. \nonumber
\end{eqnarray}
Here we also obtain only 
one fixed point $D_1^*=0$ which is stable for all temperatures. The
Lyapunov exponent is given by $\lambda_1= \frac 3 4 (1+m^2) \tanh \frac {3J} T 
+ \frac 3 4  (1-3 m^2) \tanh \frac J T -1 < 0$. Thus, the behavior is not 
chaotic and the asymptotic Hamming distance is $D=0$.
Analogously, in the ferromagnetic phase for $m^{(1)} = -m^{(2)} =m$ we obtain
a single stable fixed point $D_2^*=m$.
The results for damage spreading in the heat-bath Ising model within our
effective-field approximation are summarized in Fig. 2.
\begin{figure}
  \epsfxsize=7.0cm
  \epsfysize=7.0cm
  \centerline{\epsffile{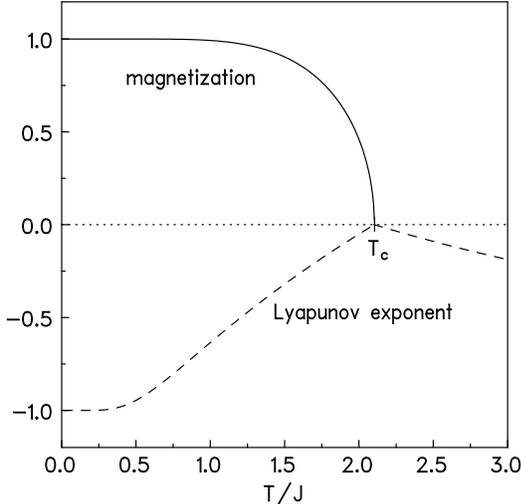}}
  \caption{Magnetization $m$ and Lyapunov exponent $\lambda_1$ as 
   functions of temperature for the heat-bath Ising model.}
  \label{fig:2}
\end{figure}

In conclusion, we studied the simultaneous time evolution of two
kinetic Ising models subjected to the same thermal noise by means of an effective 
field theory. For the 
heath-bath dynamics we found that two only slightly different equilibrium 
configurations stay close together in phase space for all times in both
the paramagnetic and the ferromagnetic phase,
i.e. an equilibrated heat-bath Ising model
does not show chaotic behavior. For the 
Glauber dynamics we found a richer behavior. For all temperatures
smaller than a spreading temperature $T_s$ the two equilibrium configurations
stay together for all times. For $T>T_s$, however, their distance
increases exponentially which corresponds to chaotic behavior. In agreement
with numerical simulations for Glauber dynamics the spreading temperature 
$T_s$ is not identical to the equilibrium critical temperature but slightly 
smaller. 

As with any mean-field theory we have, of course, to discuss in which
parameter region it correctly describes the physics of our system.
Since we treated the fluctuations in a very simplistic way, viz. treating
fluctuations at different sites as independent, our effective-field theory
will be reliable if the fluctuations are small, i.e. away from the
critical point. Therefore our theory correctly describes the high- and
low-temperature behavior whereas it might misrepresent the details close to
the critical point. In particular, the questions under which conditions
the spreading temperature coincides with the equilibrium critical
temperature and
whether the spreading transition is of 1st or 2nd order cannot be
considered solved. Further open questions are connected with the influence
of external magnetic fields, long-range interactions and disorder.
Some investigations along these lines are in progress.

This work was supported in part by the NSF under grant No. DMR-95-10185, 
and by the DFG under grant No. Vo 659/1-1.

\vfill\eject

\begin{references}
\bibitem{sym} In contrast to the Glauber dynamics the heat-bath algorithm
does not preserve the symmetry with respect to a global flip of all spins.
Therefore $D=1$ is not a fixed point here.
\end{references}
\end{document}